\documentclass[12pt]{article}

\usepackage{scicite}
\usepackage{graphicx}
\usepackage{amsmath}    
\usepackage{mathptmx}
\usepackage{amssymb}
\usepackage{times}
\makeatletter
\let\NAT@parse\undefined
\makeatother
\usepackage[
    colorlinks=true,
    linkcolor=blue,                 
    citecolor=blue,                 
    filecolor=blue,                 
    urlcolor=blue]{hyperref}        

\newenvironment{sciabstract}{%
\begin{quote} \bf}
{\end{quote}}

\renewcommand{\eqref}[1]{\hyperref[{#1}]{\textup{(\ref*{#1}})}}
\newcommand{\figref}[1]{\hyperref[{#1}]{\textbf{\textup{Figure~\ref*{#1}}}}}
\newcommand{\vect}[1]{\mathbf{#1}}          

\newcounter{lastnote}

\title{\textbf{Free electron emission in vacuum assisted by photonic time crystals}} 
\author{Xiaoke Gao, Xiaoyu Zhao, Xikui Ma, Tianyu Dong $^\ast$\\
\normalsize{School of Electrical Engineering, Xi'an Jiaotong University, Xi'an 710049, China}\\
\normalsize{$^\ast$To whom correspondence should be addressed; E-mail: tydong@mail.xjtu.edu.cn.}
}
\date{}

\begin{document} 
\baselineskip 24pt
\maketitle 

\begin{sciabstract}
The Cerenkov radiation and the Smith--Purcell effect state that free electron emission occurs exclusively in dielectrics when the velocity of the particles exceeds the speed of light in the medium or in the vicinity of periodic gratings close to each other within a vacuum. We demonstrate that free electrons in a vacuum can also emit highly directional monochromatic waves when they are in close proximity to a medium that is periodically modulated temporally, suggesting the existence of temporal Smith--Purcell effect. The momentum band gaps of time-varying media, such as photonic time crystals (PTCs), create new pathways for the injection of external energy, allowing the frequency, intensity, and spatial distribution of the electromagnetic fields to be controlled. Moreover, the PTC substrate enables the conversion of localized evanescent fields into amplified, highly directional propagating plane waves that are only sensitive to the velocity of particles and the modulation frequency, which allows us to observe and utilize Cerenkov-like radiation in free space. Our work exhibits significant opportunities for the utilization of time-varying structures in various fields, including particle identification, ultraweak signal detection, and improved radiation source design.
\end{sciabstract}

\section*{Introduction}
The investigation of interactions between charged particles and substances can date back to Cerenkov radiation \cite{cerenkov1934visible,frank1937coherent}, the Smith--Purcell effect \cite{smith1953visible,van1973smith,shibata1998coherent} and transition radiation \cite{ginzburg1979several,ginzburg1982transition,happek1991observation}, to name a few. By manipulating the properties of optical media, it is possible to obtain diversified electromagnetic radiation, thereby facilitating the development of numerous applications, such as the detection of high-energy particles \cite{marshall1952particle,saltzberg2001observation}, novel radiation sources \cite{yoshii1997radiation,liu2012surface}, free electron lasers \cite{batrakov2009carbon,song2018cherenkov}, medical imaging \cite{glaser2015cherenkov,shaffer2017utilizing}. With the emergence of metamaterials and metasurfaces, it has become possible to tailor the radiation pattern in a more comprehensive manner, including aspects such as direction \cite{genevet2015controlled,lin2018controlling}, polarization \cite{wang2016manipulating,gao2023analysis}, and intensity \cite{zhan2014tunable,liu2014special,kaminer2017spectrally}.

Conventional gain materials have generally been used to address the issue of low radiation intensity \cite{chen2023free}. However, these materials lack the ability to dynamically change the radiation direction, making them only effective as radiation sources, but insufficient for particle identification. Beyond gain materials, time-varying media have also been used to increase the intensity of electromagnetic waves in particular modes \cite{lustig2018topological,asadchy2022parametric,gaxiola2023growing}, which exhibit unique properties such as frequency conversion \cite{galiffi2020wood,wang2020theory,wang2021space,wang2023metasurface}, polarization control \cite{kalluri2018electromagnetics,akbarzadeh2018inverse, xu2021complete}, time reversal \cite{bacot2016time,vezzoli2018optical,moussa2023observation}, and energy manipulation \cite{li2021temporal,hayran2022homega}, offering significant potential for manipulating charged particle radiation patterns. In recent years, the study of the interaction between time-varying media and charged particles mainly focuses on the following two aspects: (i) the generation of radiation from stationary charges via the construction of multiple temporal interfaces \cite{li2023stationary}, similar to the concept of transition radiation at spatial interfaces \cite{lin2018controlling}; and (ii) the manipulation of radiation patterns emitted by moving charges \cite{dikopoltsev2022light} or oscillating dipoles \cite{lyubarov2022amplified} by virtue of the implementation of continuous time-varying systems and by leveraging the band structure of photonic time crystals \cite{lustig2018topological}. However, the aforementioned works involve the modification of the environment directly exposed to the particles, resulting in inevitable disruptions to the states of the particles \emph{per se}, which restricts their applications in flexible radiation sources or particle detectors. 

Instead of embedding the particles within time-varying media \cite{dikopoltsev2022light}, in our work, a rapid particle was considered in vacuum close to a photonic time crystal (PTC). Utilizing the distinctive band structures of PTC \cite{lustig2018topological} enables the conversion of electromagnetic waves from the confined modes to the propagating modes, and the parametric amplification of certain components. The proposed mechanism not only facilitates the excitation of highly directional amplified waves which is similar to the utilization of the pseudo-Brewster effect of gain materials \cite{chen2023free}, but also allows for the manipulation of propagation direction by adjusting the velocity of particles, making it suitable for both radiation sources and particle detection.

\section*{Results}
\begin{figure}[!t]    
    \centering
    \includegraphics[width=\textwidth]{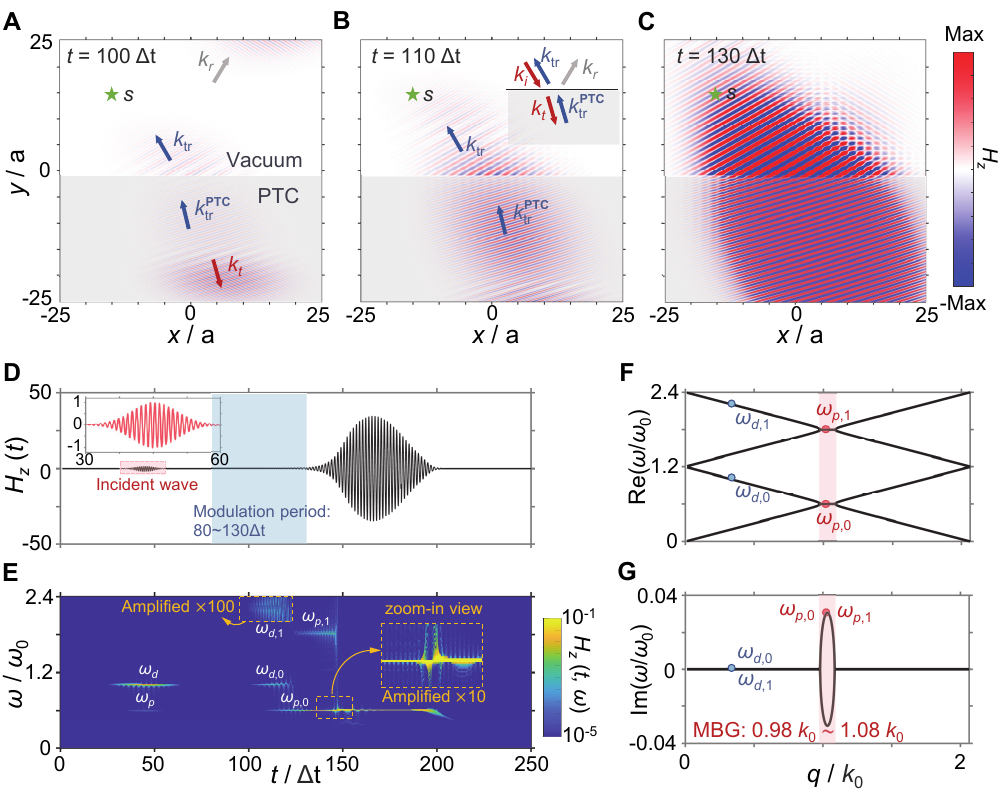}
    \caption{\textbf{Enhancement of time-reflected electromagnetic waves assisted by PTCs.} (\textbf{A}, \textbf{B} and \textbf{C}) Evolution of the magnetic field $H_z$ when a parallel-polarized Gaussian beam is incident on the vacuum--PTC interface, which correspond to $t = 100 \Delta{t}$, $t = 110 \Delta{t}$ and $t = 130 \Delta{t}$, respectively. The incident angle is $\theta_i=30^{\circ}$ with respect to the normal of the interface. The red arrows indicate the incident wave $k_i$ and transmitted waves $k_t$; the grey arrows indicate the ordinary reflected wave $k_r$; and the blue arrows indicate the extraordinary time reflected wave $k_\text{tr}$ in vacuum and $k_\text{tr}^\text{PTC}$ in the PTC. In the visualization, the values of $H_z$ in the PTC of the lower half-space have been reduced by $1/5$ to emphasize the field in the free space of the upper half-space. (\textbf{D}) The waveform in the time domain of $H_z$ at the point $s=(-15 a, 15 a)$ marked by a green star in \textbf{A}. The inset illustrates a zoom-in view of the incident wave within the red box, and the blue shaded region demonstrates the modulation period of the PTC. (\textbf{E}) Results of time-frequency analysis for \textbf{D} using the Synchrosqueezed Wavelet Transform (SSWT). (\textbf{F} and \textbf{G}) The dispersion relation for the extraordinary waves denoted by $k_\text{tr}^\text{PTC}$ in the PTC; \textbf{F} displays the real component of the frequency $\omega$, and \textbf{G} depicts the imaginary part. The shaded region illustrates the momentum band gap ranging from $0.98 k_0$ to $1.08 k_0$. In addition, the corresponding modes for the TR wave are labeled in markers.}
    \label{fig:figure01}
\end{figure}

\paragraph{Exponential amplification of electromagnetic waves by time-varying media.} 
Materials undergoing periodically temporal changes exhibit momentum band gaps (MBGs), allowing for non-conventional energy injection or amplification of ultra-weak disturbances by facilitating exponential amplification of particular electromagnetic modes. We begin by investigating the interaction between a Gaussian beam (plane wave) and a planar vacuum--PTC interface, which has a physical interpretation similar to scenarios involving moving particles. \figref{fig:figure01}A -- \figref{fig:figure01}C illustrate the evolution of electromagnetic waves during the modulation of the PTC. In addition to ordinary incident and reflected waves in vacuum, extraordinary electromagnetic waves appear when the PTC is present, which are referred to as time-reflected (TR) waves, exhibiting both inside the PTC and within the vacuum, and traveling in the opposite direction of the incident waves. Moreover, during modulation, the intensity of the time-reflected wave would increase exponentially with time, suggesting its potential use in distinguishing weak signals. \figref{fig:figure01}D plots the time-domain waveform of the magnetic field $H_z$ at the sampling point $s$ (labeled as green stars in \figref{fig:figure01}A -- \figref{fig:figure01}C). When $30 \Delta{t} < t < 60 \Delta{t}$, the incident modulated Gaussian sine wave arrives at the sampling point $s$, whose dominant component has a magnitude of one with a frequency of $\omega_d = \omega_0$ and the perturbed component has a magnitude of 0.02 with $\omega_p = 0.6\omega_0$ (Here, $\Delta{t}$ is the time step in the simulation and $\omega_0 = 2\pi$ is used throughout; see \textbf{Materials and Methods} for details). Unlike the conventional interface with pure spatial discontinuity, for which no reflected wave can be observed at the sampling point $s$, the reflected waves present at $s$, which have a significantly higher intensity than the incident wave. \figref{fig:figure01}E displays the time-frequency analysis of the waveform shown in \figref{fig:figure01}D. Within the PTC modulation, extraordinary TR waves appear with frequencies $\omega_d + n\Omega$ and $\omega_p + n\Omega$, which is consistent with the Floquet theorem \cite{lustig2018topological,wang2023metasurface}. Furthermore, only the modes of $n = 0$ and $n = 1$ are retained, because the magnitude of the other higher modes becomes very small. \figref{fig:figure01}F and \figref{fig:figure01}G depict the dispersion relation of the PTC, indicating that complex frequency modes could exist within the MBG. In MBG, $\Re[\omega] = (n+1/2)\Omega$ \cite{park2021spatiotemporal,asadchy2022parametric}; however, the imaginary parts exhibit non-zero values and bifurcate: one corresponds to an exponential growing wave, and the other one denotes an exponential decay in time. Outside the MBG, the eigenfrequencies are pure-real, and the corresponding modes would not be amplified. Therefore, although the perturbed component of the incident wave with frequency $\omega_p = \Omega/2$ has a much smaller magnitude compared to the dominant component of frequency $\omega_d \neq \Omega/2$, it would be significantly amplified, resulting in high-magnitude modes with frequencies $\omega_{p,n} = (n+1/2)\Omega$, \emph{e.g.}, the modes denoted by $\omega_{p,0}$ and $\omega_{p,1}$ in \figref{fig:figure01}E and \figref{fig:figure01}F. We stress that the amplification of waves within a PTC occurs during a specific time period, as opposed to conventional gain materials, in which this is realized as the waves propagate over a distance facilitated by complex wave vectors. This characteristic suggests that the proposed technology has the potential to function as a space-compact amplifier.

Within a PTC substrate, incident waves of frequency $\omega_p = \Omega/2$ can be amplified and produce TR waves propagated in the opposite direction, which are characterized by frequencies $\omega_{p,n} = (n+1/2)\Omega$. For a moving particle in vacuum near the PTC, spatial decaying waves in the free space with a continuous spectrum radiate; therefore, the components of the evanescent waves with frequency $\omega_p$ can be amplified and converted into propagating waves with frequencies $\omega_{p,n} = (n+1/2)\Omega$, which is beneficial for detecting the ultraweak signals introduced by moving particles. Due to the conservation of tangential components of the wave vectors at the vacuum--PTC interface, when charges move in parallel to the interface, it is possible to identify particles of various velocity $v$ by determining the tangential component of the wave vector $k_x = k_0 \cos\theta = \omega/c_0 \cos\theta$ and the radiation frequency $\omega$ of the propagating waves in the free space, \emph{i.e.}, $v = \omega/k_x$, where $\theta$ denotes the radiation angle with respect to the interface.

\begin{figure}[!htb]    
    \centering
    \includegraphics[width=0.90\textwidth]{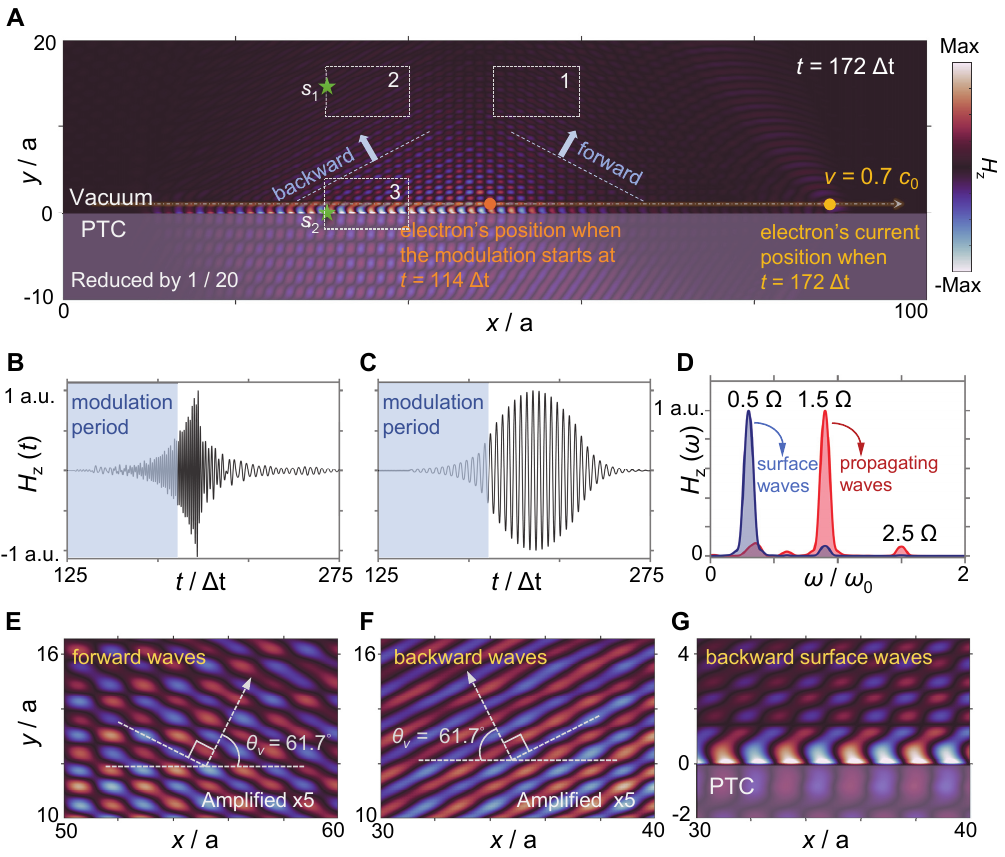}
    \caption{\textbf{Enhanced free electron radiation in vacuum assisted by PTCs.} (\textbf{A}) Near-field distribution when $t=172 \Delta{t}$. Here, the trajectory (yellow dashed line) of the swift electron is along the $x$-axis, which is parallel to the vacuum--PTC interface located at $y = 0$. Electrons start to move at a velocity $v = \beta \text{c}_0$ when $t = 0$, where $\beta$ is the ratio between the velocity of the particle and the light speed $c_0$ in vacuum; and the modulation period is $114 \Delta{t} \leq t \leq 181 \Delta{t}$. The yellow and orange circles indicate the positions of the particles when $t = 172 \Delta{t}$ and $t = 114 \Delta{t}$, respectively; the blue arrows indicate the direction of the plane wave. In the visualization, the values of $H_z$ in the PTC have been reduced by $1/20$ to emphasize the field in the free space. (\textbf{B}) and (\textbf{C}) Normalized time-domain waveforms $H_z$ of the backward plane waves and the surface waves at the points $s_1$ $(30 a, 15 a)$ and $s_2$ $(30 a, 0)$, respectively, marked by green stars in \textbf{A}. The modulation period of the PTC is shaded in blue. (\textbf{D}) Frequency analysis of the time-domain waveforms of \textbf{B} and \textbf{C}, which correspond to the red and blue curves, respectively, with the fraction of waveforms during $170 \Delta{t} < t < 195 \Delta{t}$ and $155 \Delta{t} < t < 180 \Delta{t}$ being used in the Fast Fourier Transform (FFT). (\textbf{E}), (\textbf{F}) and (\textbf{G}) Zoom-in views of the magnetic field $H_z$ for forward propagating waves, backward propagating waves, and surface waves, respectively, within boxes 1, 2 and 3 in \textbf{A}. In \textbf{E} and \textbf{F}, the values have been amplified by five to emphasize the wave behavior. The forward and backward waves have the same radiation angle $\theta_v = 61.7^\circ$.}
    \label{fig:figure02}
\end{figure}

\paragraph{Enhanced free electron emission in vacuum.}
Now, we consider the details of the scenario in which free electron beams move in vacuum near the PTC, as depicted in \figref{fig:figure02}A. \figref{fig:figure02}A shows the near-field distribution when $t = 172 \Delta t$. Remarkably, propagating waves can be observed in the free space in both the forward and the backward directions, which is analogous to the traditional Smith--Purcell radiation. Furthermore, the waves generated are highly directional, which is absent in Smith--Purcell radiation \cite{donohue2005simulation, li2006particle}. In addition, high-intensity surface waves exist at the interface due to the total internal reflection of components in TR waves within the PTC. \figref{fig:figure02}B and \figref{fig:figure02}C demonstrate the time-domain waveforms of the backward propagating waves and the backward surface waves at the sampling points $s_1$ and $s_2$, respectively. During PTC modulation, the amplitudes of the propagating and surface waves would grow exponentially. \figref{fig:figure02}D illustrates the spectrum of the waveforms depicted in \figref{fig:figure02}B and \figref{fig:figure02}C. For the surface wave, the dominant frequency is $\omega_{s,0} = \Omega/2$ due to total reflection and the other components with frequencies $\omega_{s,n} = (n+1/2)\Omega$ are small when $n \geq 0$. For propagating waves, the dominant component has frequency $\omega_{p,1} = 3\Omega/2$ and the component with frequency $\omega_{p,0} = \Omega/2$ is suppressed because it behaves as surface waves. Here, only the modes of $n = 0, 1, 2$ are retained because the other higher modes become very weak. \figref{fig:figure02}E -- \figref{fig:figure02}G exhibit zoom-in views of the magnetic field distribution of the backward and forward propagating waves, and the surface waves, respectively, as illustrated by the dashed boxes in \figref{fig:figure02}A. It is evident that the surface waves have a much stronger magnitude than the propagating waves. According to Fourier analysis, the tangential component of wave vectors for forward/backward propagating waves and surface waves are identical and equal to $k_x = 0.428 k_0$, which can be calculated by $k_x = \Omega/(2\beta c_0)$ with $\beta = 0.7$ and $k_0 = \omega_0/c_0$. Furthermore, the radiation angles of the forward and backward waves measured in \figref{fig:figure02}E and \figref{fig:figure02}F are identical and read $\theta_v = 61.7^{\circ}$, which is consistent with $\theta_v = \arccos[(2n+1)^{-1}\beta^{-1}]$ for $n = 1$. In summary, the frequency $\omega$ and the radiation angle $\theta$ of the propagating waves are determined by the modulation frequency $\Omega$ of the PTC and the charge velocity $\beta$, respectively, \emph{i.e.}, $\omega = 3 \Omega / 2$ and $\theta = \arccos 1/(3\beta)$, resulting in a completely new mechanism for particle identification. Furthermore, the intensity of propagating waves and surface waves can increase exponentially during modulation (see S3 in the \textbf{Supplementary Material} for the snapshots of field distribution at various time), allowing waves to be detected more easily compared to traditional Cerenkov radiation \cite{shaffer2017utilizing}, Smith--Purcell effect \cite{liu2014special}, and transition radiation \cite{ginzburg1982transition, lin2018controlling}.

\begin{figure}[!htb]    
    \centering
    \includegraphics[width=\textwidth]{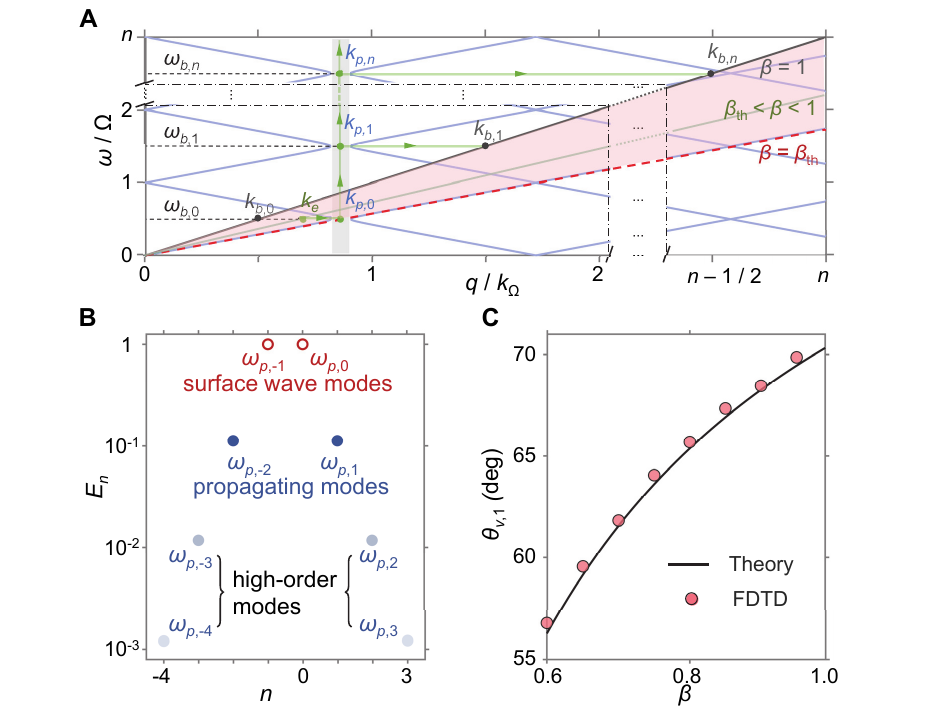}
    \caption{\textbf{Mechanism of the conversion between propagating and confined modes.} (\textbf{A}) Band structure of the PTC whose relative permittivity reads $\epsilon(t) = \epsilon_r ( 1 + \alpha \sin\Omega t)$ with $\epsilon_r = 3$ and $\alpha = 0.2$. The dispersion curves of the PTC are depicted in blue; the black line represents the light line $\beta = 1$ in vacuum; the red dashed line represents the threshold $\beta = \beta_\text{th}$, where $\beta_\text{th} = 1/\sqrt{\epsilon_r} \approx 0.58$ in the PTC; the green line represents the dispersion relation for a confined wave near the electron. The grey-shaded region illustrates the MBG when $0.82 k_\Omega < q < 0.9 k_\Omega$, where $k_\Omega = \Omega/c_0$. The red-shaded region illustrates the range for which confined modes can be converted into propagating modes. The path of mode evolution is illustrated by the green lines with arrows. (\textbf{B}) Normalized amplitudes of the elements $\vect{E}_n$ of the eigenvector $\Phi$ corresponding to $\omega_{p,n}$ for the eigenvalues $q=k_{p,n}$ within the MBG. Here, the maximum mode number $N$ is truncated to $N=4$. The surface wave modes with frequencies $\omega_{p,-1}$ and $\omega_{p,0}$ are labeled in hollow red circles. The other modes are propagating modes, which are labeled by solid blue circles; and the darker color indicates greater amplitudes. (\textbf{C}) Theoretical and numerical results of the radiation angle $\theta_{v,1}$ with respect to the velocity of the particles $\beta = v/c_0$, for which the theoretical results are calculated by \eqref{eq:thetavn} and the numerical results are obtained using the finite difference time domain (FDTD) method with the open source MEEP software package. }
    \label{fig:figure03}
\end{figure}

\paragraph{Conversion between the propagating and the confined modes.} 
\figref{fig:figure03}A displays the band structure of the PTC for \figref{fig:figure02} whose relative permittivity reads $\epsilon(t) = \epsilon_r (1+\alpha\sin\Omega t)$ with $\epsilon_r = 3$ and $\alpha = 0.2$, which can be calculated using the Floquet theorem \cite{stefanou2021light} (see S2 in the \textbf{Supplementary Material} for the details). The plane waves propagating in vacuum with frequency $\omega_{b,0} = \Omega/2$ and wave number $k_{b,0} = \omega_{b,0}/c_0$ correspond to the operating point $k_{b,0}$ on the grey light line for $\beta = 1$ in \figref{fig:figure03}A. For particles traveling at a velocity of $\beta c_0$ in vacuum, electromagnetic waves of frequency $\omega_{b,0}$ and wave number $k_e = \omega_{b,0}/(\beta c_0)$ may be produced, corresponding to the operating point $k_e$ on the green line for $\beta$ in \figref{fig:figure03}A. Since $\beta < 1$ for any particles, the electromagnetic fields induced by the particle are confined to the particle itself because $k_e > k_{b,0}$ and would not propagate. However, when such a confined field is coupled to the PTC, waves with various frequencies $\omega_{p,n}$ can be excited within the PTC, which have the same wave number $k_{p,n}$ in the PTC. As discussed previously, the excited modes are located within the MBG of the PTC, exhibiting complex frequencies $\omega_{p,n}$ such that $\Re[\omega_{p,n}] = \omega_{b,n} = \omega_{b,0} + n\Omega$ and identical transverse wave number $k_x = k_e = \omega_{b,0} / (\beta c_0)$ because the tangential wave vectors must be continuous at the spatial discontinuity of the vacuum--PTC interface. Interestingly, modes with frequencies $\omega_{p,n}$ when $n \geq 1$ may have a chance to escape the PTC region to the neighboring free space since $k_{b,n} = \omega_{b,n}/c_0 > k_x$. In \figref{fig:figure03}A, the green lines with arrows illustrate the evolution from confined fields labeled $k_{b,0}$ to propagating waves labeled $k_{b,n}$ with frequencies $\omega_{b,n}$ and wave numbers $k_{b,n}$. As a result, the radiation angle $\theta_{v,n}$ for the $n$-th order propagating waves can be determined by the electron velocity $\beta c_0$ and the mode number $n$, which reads
\begin{equation} \label{eq:thetavn}
    \cos{\theta_{v,n}} = \frac{k_x}{k_{b,n}} = \frac{\omega_{b,0}/(\beta c_0)}{\omega_{b,n}/c_0} = \frac{1}{\beta} \frac{1}{2n+1}.
\end{equation}
Alternatively, considering $\omega_{b,n} = \omega_{b,0} + n \Omega$ and $\omega_{b,0} = 2\pi c_0 / \lambda_{b,0}$, \eqref{eq:thetavn} can be rewritten as $\lambda_{b,0} = \frac{d_{\Omega}}{n} \left( \frac{1}{\beta \cos\theta_{v,n}} - 1 \right)$, where $\lambda_{b,0}$ denotes the radiated wavelength and $d_{\Omega} = 2\pi c_0 / \Omega$, which resembles the relationship between the radiated wavelength, the pitch of the grating, the velocity of the particles, and the angle of radiation in the conventional Smith--Purcell effect \cite{andrews2005dispersion}. This implies that the PTC-assisted free-electron emission in vacuum may be interpreted as a temporal analogy to the classical Smith--Purcell effect.

\figref{fig:figure03}B displays the normalized amplitudes of the elements of the eigenvector $\Phi_n$ ($-N\leq n \leq N$) corresponding to $\omega_{p,n}$ for the eigenvalues $q = k_{p,n} = 0.8660 k_\Omega$ in the MBG, where $k_\Omega = \Omega/c_0$. Here, the mode number $n$ is truncated to $N = 4$. For the modes $n = -1$ and $n = 0$ with frequency $\omega_{b,0} = \Omega / 2$ that correspond to surface waves, although the intensities are the greatest, it may face challenges in practical application due to measurement issues. In contrast, for the modes $n = -2$ and $n = 1$ whose frequency reads $\omega_{b,1} = \omega_{b,0} + \Omega = 3\Omega/2$, even the corresponding eigenmode amplitude is not the greatest, they can be amplified due to the MBG. In addition, other higher modes are too weak to be contributed. Therefore, only the modes of $n = 0, 1$ are retained in \figref{fig:figure03}A. For the fundamental mode when $n = 0$, we have $\cos\theta_{v,0} = 1/\beta > 1$, indicating total reflection and producing surface waves. When $\beta_\text{th} \leq \beta < 1$, although $k_e > \omega_{b,0}/c_0$, by increasing the frequency from $\omega_{b,0}$ to the next modes of frequencies $\omega_{b,n} = \omega_{b,0} + n\Omega$ within the MBG such that $k_{b,n} > k_x = k_e$, allowing the longitudinal wave number $k_y$ to be real and the waves to propagate in the free space. Moreover, since only the mode with frequency $\omega_{b,1}$ is dominant (see also \figref{fig:figure03}B), highly directional monochromatic plane waves can be observed in vacuum, whose frequency and wave number are $\omega_{b,1} = 3\Omega/2$ and $k_{b,1} = \omega_{b,1}/c_0$, respectively. As a result, the angle of the plane wave radiation can be expressed as $\theta_v = \theta_{v,1} =  1/(3\beta)$. \figref{fig:figure03}C illustrates the relation between the velocity of the particle $\beta c_0$ and the radiation angle $\theta_{v,1}$, demonstrating good agreement with the simulation results. When $\beta < \beta_\text{th}$, the tangential component of the wave vector $k_e$ of the confined mode corresponding to frequency $\omega_{b,0}$ exceeds the MBG of the PTC; consequently, the inter-band mode in the PTC cannot be excited and high-intensity radiation in vacuum no longer exists (see S3 in the \textbf{Supplementary Material} for details).

If the static permittivity $\epsilon_r$ is large and $\beta_\text{th} = 1/\sqrt{\epsilon_r}$ is small, the particle velocity allowed $\beta$ can be small such that $ (2n + 1)\beta < 1$ if $n$ is small; therefore, \eqref{eq:thetavn} may not be valid for small mode numbers. When $\beta > \beta_\text{th}$, there exists a critical integer $n_c \in \mathbb{Z}$ such that $\beta_\text{th} < \beta < 1/(2n_c + 1)$, implicating particles traveling with velocity $\beta$ can produce propagating modes in vacuum of order $n > n_c$. As $\beta$ decreases for low-energy particles, the corresponding tangential component of the wave vector $k_e$ increases, requiring a higher frequency $\omega + n\Omega$ to ensure that the normal component of the wave vectors of the propagating waves is real, \emph{i.e.}, $(\omega + n\Omega) / c_0 > k_e$. This guarantees the conversion from the confined modes to the propagating modes (see S3 in the \textbf{Supplementary Material} for details). Now, the mode number of the excited wave propagating at the lowest frequency reads $n_p = \lceil (1/\beta - 1)/2 \rceil > 1$. Similarly to the scenario where $n_p = 1$, the other higher modes for $n > n_p$ would be weak and negligible. As a result, highly directional monochromatic propagating waves of frequency $\omega_{b,n} = (n_p + 1/2)\Omega$ may also be detected, whose radiation angle is determined by \eqref{eq:thetavn} where $n = n_p$.

\section*{Discussion}
In summary, we have revealed an unprecedented physical mechanism of the interaction between moving particles and photonic time crystals, which allows free electrons in vacuum to radiate highly directional monochromatic plane waves in the free space. In akin to traditional Smith--Purcell radiation, for which spatial periodicity is introduced to compensate for transverse momentum, along the interface, the frequency of nonradiative waves can be increased within the temporal periodicity so that the corresponding wave numbers in the free space are greater than the transverse momentum, which would result in propagating waves in vacuum, since the spatial translation symmetry ensures the conservation of the components of the wave vector along the trajectory of particles. Moreover, the intensity of the propagating wave can be exponentially amplified due to the MBG of the PTC during modulation. Remarkably, this enhanced radiation is highly directional and monochromatic, allowing us to determine the velocity of the particles by measuring the radiation angle. These advantages show promise for a wide range of applications, including the detection of ultraweak signals and the development of advanced radiation sources, compared to traditional Cerenkov radiation and the Smith--Purcell effect. From a practical application perspective, the radiation angle is only determined by the velocity of the particles and would not be affected by material parameters such as modulation depth, period, and duration. Moreover, the frequency of the propagating waves is exclusively subject to the modulation frequency, suggesting its potential usage as a monochromatic radiation source. In addition, the mechanism is valid for any modulation frequency, showing a broad spectrum of possible applications in the RF/THz regimes. Our study demonstrates that waves can be manipulated from the perspective of energy and momentum by incorporating temporal modulation, allowing both the frequency and the wave vector to be controlled concurrently.

\section*{Materials and Methods}
The plane wave expansion method can be used to calculate the dispersion relation (band structures) of a material system and the corresponding eigenstates or eigenmodes \cite{joannopoulos2008photonic} (see S2 in the \textbf{Supplementary Material} for the details). The magnetic field distributions were calculated using an open source software package MEEP \cite{oskooi2010meep}, which is based on the finite difference time domain (FDTD) method. By considering the scale-invariance of Maxwell's equations \cite{joannopoulos2008photonic}, scale-invariant units are employed for which the frequency $\omega_0 = 2\pi$ and $a = 1$. The relative permittivity of PTC is $\varepsilon(t) = \epsilon_r(1+\alpha\sin\Omega t)$ with $\epsilon_r$ being the static permittivity, $\alpha$ and $\Omega$ being the modulation depth and frequency, respectively, which was modeled by modifying the permittivity at each time step. The parameters are $\epsilon_r = 3$ and $\alpha = 0.2$ throughout the simulation; and $\Omega = 1.2 \omega_0$ and $\Omega = 0.6 \omega_0$ are adopted respectively for the illustrations of plane wave reflection and refraction (\figref{fig:figure01}), and free electron emission (\figref{fig:figure02}, \figref{fig:figure03}). For plane waves, two $p$-polarized modulated Gaussian beams with different central frequencies are superposed at the top boundary, which have the same incident angle $\theta_i = 30^{\circ}$ and waist radius $W_0 = 8a$. However, their amplitudes of the form $H^\text{inc} = \exp[-(t-t_c)^2/\tau_c] H_0 \cos{\omega_c (t-t_c)}$ have different parameters, which read $t_c = 0$, $\tau_c = 10\pi / \omega_0$, $H_0 = 1$, and $\omega_c = \omega_0$ for the dominant component and $t_c = 0$, $\tau_c = 10\pi / \omega_0$, $H_0 = 0.02$, and $\omega_c = 0.6\omega_0$ for the perturbed component. The moving electron was simulated by updating the position of the electric dipole at each time step. In the post-processing visualization, the field distribution in PTC is reduced several times to emphasize the field or wave behavior in vacuum, as described in the figure captions. Detailed information on the simulation is provided in S4 of the \textbf{Supplementary Material}.

\section*{Acknowledgements}
T.D. acknowledges the support of the National Natural Science Foundation of China (NSFC) under grant no. 51977165. X.G. thanks the support from the Fundamental Research Funds for the Central Universities (xzy022023035).

\section*{Author Contributions}
All authors contributed to all aspects of this work.

\section*{Data Availability}
All key data supporting the findings are included in the main text and its supplementary material. Additional data sets are available from the corresponding author upon reasonable request.

\section*{Code Availability}
The source codes and simulation files that support the figures and data analysis in this article are available from the corresponding author upon reasonable request.

\section*{Competing Interests}
The authors declare that they have no competing interests.

\baselineskip 12pt
\bibliography{Main}
\bibliographystyle{Science}

\vspace{2em}
\begin{flushright}
Dated: November 2, 2023 \\
\end{flushright}

\clearpage

\clearpage

\clearpage

\end{document}